\documentclass[aps,prl,tightenlines,twocolumn,superscriptaddress]{revtex4}
\usepackage{amssymb}
\usepackage{color,graphicx}
\usepackage{enumitem}
\usepackage{amsmath}
\usepackage{amsbsy}
\usepackage{amsthm}
\usepackage{bbm}
\usepackage{bm}
\usepackage{epsfig}
\setdescription{leftmargin=\parindent,labelindent=\parindent}

\begin{document}

\title{On the Meaning of Locality: The Overlapping Assumptions}

\author{Afshin Shafiee}
\email[Corresponding Author:~]{shafiee@sharif.edu}
\affiliation{Research Group On Foundations of Quantum Theory and Information, Department of Chemistry, Sharif University of Technology P.O.Box 11365-9516, Tehran, Iran}
\affiliation{Foundations of Physics Group, School of Physics, Institute for Research in Fundamental Sciences (IPM), P.O. Box 19395-5531, Tehran, Iran}

\author{Farhad Taher Ghahramani}
\email[]{farhadtqm@ipm.ir}
\affiliation{Foundations of Physics Group, School of Physics, Institute for Research in Fundamental Sciences (IPM), P.O. Box 19395-5531, Tehran, Iran}

\begin{abstract}
We examine the Locality assumption of Bell's theorem in three steps of EPRB experiment. Depending on the context, Locality is embodied in the conditions of Separability, Local Causality, Factorizability, Relativistic Causality, and Non-Contextuality. We show that Separability, characterized by the constraint of zero covariance, is equivalent to Local Causality which leads to Factorizability through the measurement process. Factorizability is the conjunction of Parameter Independence and Outcome Independence. It is commonly believed that Relativistic Causality is equivalent to Parameter Independence, which is satisfied by Quantum Mechanics. According to our approach, however, this is unjustified due to the second step analysis of EPRB experiment, including the measurement on the first particle. We define Non-Contextuality based on preparation procedure and measurement process. Preparation-based non-contextuality is an independent assumption, but non-locality within the framework of a separable model can be interpreted as measurement-based contextuality. Finally, we show that any fundamental theory consistent with Quantum Mechanics, should refute Outcome Independence in its framework of description.
\end{abstract}

\maketitle

\section{1. Introduction}
Bell's theorem is one of the most profound discoveries of science. It provides experimentally-testable predictions about the metaphysical assumptions. Shimony coined the term ``experimental metaphysics'' to point out the experimental status of metaphysical assumptions of Bell's theorem~\cite{Shi}. The problem is originally traced back to the EPR argument~\cite{EPR}. EPR showed that if we assume locality principle is respected, the predictions of Quantum Mechanics (QM) will lead to pre-existing values of position and momentum for previously-interacted particles. Since orthodox QM denies the existence of such pre-existing outcomes, EPR claimed to establish the incompleteness of QM. The argument was later reformulated by Bohm in terms of spin~\cite{Boh1}. This ``EPRB'' version is conceptually simpler and also more closely related to the recent experiments designed to examine Bell's theorem. The structure of the original EPR argument transcribed into Bohm version is as follows: consider two spin-half particles prepared in a singlet state. QM predicts that the outcomes of measurements of the spin of two particles along any axis are perfectly anti-correlated. If such measurements are carried out at space-like separation, then locality principle requires that the outcome of the measurement on one particle is independent of the measurement on the other particle. So, the only way to realize the perfect anti-correlation is to assume pre-existing values for the measured properties. \\
\indent Along with EPR, a different, but closely related program was triggered by von Neumann~\cite{Neu}, and followed by others~\cite{Jau,Gle} trying to show the impossibility of any deterministic reconstruction of QM by incorporating additional parameters, known as hidden variables (HVs). However, Bohm's deterministic model~\cite{Boh2}, better known as Bohmian Mechanics (BM), was the demonstration of the faultiness of the impossibility theorems~\cite{Bel1}. All these theorems were based on an assumption, known as non-contextuality. A theory that assigns well-defined values to all quantum observables at all times is usually called a non-contextual HV theory. The impossibility of such a theory follows from Bell's theorem~\cite{Bel2}, and more explicitly from Kochen-Specker (KS) theorem~\cite{Koc}. What do we exactly mean by the term ``non-contextual'', however? Consider observables $A$, $B$ and $C$ with $[A,B]=[A,C]=0$, but $[B,C]\neq0$. Suppose that $A$ is measured along with $B$ or along with $C$. If one assumes that such measurements reveal pre-existing values, the outcomes of the measurement of $A$ should be the same for both experiments. However, since the observables $B$ and $C$ are not compatible, these experiments are different. More precisely, within a contextual HV theory, the outcomes of experiments are determined by two different kinds of HVs for two different experimental contexts. In examining how BM managed to violate non-contextuality, Bell noticed that the theory is manifestly non-local~\cite{Bel1}. At this point, two seemingly different programs intertwined. \\
\indent Subsequently, Bell showed that deterministic HVs resulted from the locality assumption in the EPR argument satisfy an inequality incompatible with the predictions of QM~\cite{Bel2}. Whereas the original Bell's theorem deals only with deterministic HVs based on the perfect anti-correlations, the theorem was later generalized to stochastic theories as well. The results are known as CHSH~\cite{CHSH,Bel3} and CH~\cite{CH} inequalities. Also, some approaches have been suggested based on perfect correlations instead of probabilistic inequalities~\cite{Hem,GHZ,Har}. The contradiction between these generalized Bell inequalities and the predictions of QM could be examined by experiment. However, there might be problems in experimental set-ups, often referred to as ``loopholes'', that affect the validity of the experimental results. Nevertheless, the series of increasingly sophisticated experiments have convinced the physics community that at least some special forms of the Bell inequalities are experimentally violated. So, it appeared that breakdown of locality principle, alias non-locality, is a crucial fact at the microphysics level. \\
\indent Despite the exact algebraic derivations of non-local theorems including Bell's theorem and recently Leggett's one~\cite{Leg}, there continues to be controversy whether what type of a fundamental theory, these theorems deny, so that it could be consistent with QM. The answer might be sought partly by focusing on the assumptions the theorems are contingent upon. Although great progress has been recently achieved in clearing up the crucial assumptions involved, there are conceptual gaps left open between the questions and the answers provided.\\
\indent The structure of the article is as follows. In the second section, we critically review the current ideas about the Locality assumption in the Bell's theorem to reach some consistent criteria to describe it. In the light of these criteria, in the third section, three steps of EPRB experiment are proposed and analyzed with the aim of addressing the above questions. We sum up the results in the conclusion.
\section{2. Locality Assumption}
Assuming the exclusion of loopholes to explain observed experimental violations, the central problem concluded from Bell's theorem is about the nature of non-local phenomena. It is usually believed that the development of a typical non-local theorem relies on some additional assumptions as well. Here, we are trying to clear up the meaning of locality and its distinction from the overlapping concepts.\\
\indent To grasp the core idea of the issue, let us begin with Bell's own definition of locality, better known as Local Causality (LC)~\cite{Bel4}
\begin{description}
   \item[Local Causality]{\it A theory is locally causal iff the events related to local beables in space-like separated regions are independent, given the specification of sufficient local beables in the past light cone of each event.}
\end{description}
For a more in-depth analysis of Bell's adaptation of locality principle see~\cite{Nor2}. To proceed, we need to introduce a concrete experimental set-up. The most common setting used in relevant discussions involves measurements on two particles emitted by a source, where each particle is measured by a different party, named as usual, Alice and Bob. For a certain state of entangled particles, Alice (Bob) performs measurements on the particle $1$ ($2$) along the measurement setting $\hat a$ ($\hat b$) to obtain outcome $A$ ($B$). The probability distribution $P(A,B|a,b)$ obtained in many trials of an experiment may actually arise out of a statistical mixture of HVs collectively labelled by $\lambda$ as
\begin{equation}
\label{Eq1}
P(A,B|a,b)=\int_{\Lambda} d\lambda~\rho(\lambda)P(A,B|a,b,\lambda)
\end{equation}
where $\rho(\lambda)$ is the probability distribution of $\lambda$ over the space $\Lambda$. What do we mean by $\lambda$ exactly? It indicates the specification of the state of the two-particle system under study by the relevant beables in the past light cone of each particle. This specification depends on the model under study. It may include the information of the wave function along with some additional information specified by HVs. It is incorporated into the dynamics for revitalizing certain features which are absent in QM. The most important features are determinism, ruled out by impossibility theorems, and locality, invalidated by Bell's theorem. The only constraint on $\lambda$ is the condition of Measurement Independence~\cite{Bel4}
\begin{description}
   \item[Measurement Independence]{\it The choice of measurement settings $a$ and $b$ can be freely made, independent of $\lambda$.}
\end{description}
This is a natural requirement, which is often thought of as expressing the ``free will'' of the experimenters. The violation of this condition can manifest models reproducing the predictions of QM~\cite{SHC,Bran,LPJ}. The amount of the violation, necessary to reproduce the predictions of QM was recently quantified~\cite{LPJ,Bar}. However, since this violation implies that there is some incredible and seemingly improbable conspiracy of nature, we take it as unacceptable. Note that contrary to measurement settings which are subjected to the condition of Measurement Independence, the distant outcomes can be {\it implicitly} contained in the shared region of information of $\lambda$.\\
\indent We will now determine what the definition of LC implies to our experimental situation. The definition obviously involves no measurement process. If we incorporate the measurement process in it, we arrive at the conclusion that an underlying theory satisfies LC iff
\begin{align}\label{Eq2}
P(A|a,b,B,\lambda)&=P(A|a,\lambda), \nonumber \\
P(B|a,b,A,\lambda)&=P(B|b,\lambda)
\end{align}
Using Bayes' theorem,
\begin{equation}\label{Eq3}
P(A,B|a,b,\lambda)=P(A|a,b,B,\lambda)P(B|a,b,\lambda)
\end{equation}
we conclude that LC leads to Factorizability, i.e.,
\begin{equation}\label{Eq4}
P(A,B|a,b,\lambda)=P(A|a,\lambda)P(B|b,\lambda)
\end{equation}
also known as ``Bell's locality condition''. This is the basic formulation of locality principle in non-local theorems. In an important analysis, Jarrett pointed out that Bell's locality condition is a logical conjunction of two independent conditions named after by Shimony as Outcome Independence (OI), i.e., the {\it explicit} independence of one's outcome from the other's outcome
\begin{align}\label{Eq5}
P(A|a,b,B,\lambda)&=P(A|a,b,\lambda), \nonumber \\
P(B|a,b,A,\lambda)&=P(B|a,b,\lambda)
\end{align}
and Parameter Independence (PI), i.e., the independence of one's outcome from the other's parameter~\cite{Jar,Shi1}
\begin{align}\label{Eq6}
  P(A|a,b,\lambda)&= P(A|a,\lambda), \nonumber \\
  P(B|a,b,\lambda)&= P(B|b,\lambda)
\end{align}
So, the violation of PI and/or OI leads to the violation of Bell's locality condition. The choice depends on the model under study. Some models respect both, as Bell's model. Some respect PI, but not OI, as QM. Some respect OI, but not PI, as BM (since determinism, i.e., $P(A,B|a,b,\lambda)\in \{0,1\}$, is strictly stronger than OI, any deterministic model respects OI). Some respect neither, as Nelson's stochastic mechanics~\cite{Nel1,Nel2}.\\
\indent Depending on the context, there are different adaptations of Bell's locality condition. Although they are overlapping, and often commentators use them interchangeably, each relies on a specific trait of locality principle. Here, we carefully examine these adaptations.
\subsubsection{Locality vs. Relativistic Causality}
The most relevant belief is that locality is supported by Relativistic Causality (RC), a condition motivated by Special Relativity. RC denotes the impossibility of superluminal propagation of causal influences. There is no consensus on the physical realizations of causal influences. Jarrett takes them as signals, and as a consequence, RC as No-Signalling theorem, i.e., the impossibility of superluminal signalling, which is characterized by~\cite{Pop}
\begin{align}\label{Eq7}
  P(A|a,b)&=\int_{\Lambda} d\lambda~\rho(\lambda)P(A|a,b,\lambda) \nonumber \\
  &=\int_{\Lambda} d\lambda~\rho(\lambda)P(A|a,\lambda)=P(A|a), \nonumber \\
 P(B|a,b)&=\int_{\Lambda} d\lambda~\rho(\lambda)P(B|a,b,\lambda) \nonumber \\
  &=\int_{\Lambda} d\lambda~\rho(\lambda)P(B|b,\lambda)=P(B|b)
\end{align}
\noindent Jarrett argues that since PI is a direct consequence of RC and therefore ought to be maintained, OI is the condition to be blamed. Contrary to PI, non-locality due to the violation of OI is uncontrollable and therefore does not permit superluminal signalling. It is extensively argued that this is not an instance of ``action at a distance'' but only of some innocent ``passion at a distance''. So, Jarrett suggested that Bell's locality condition is a ``strong'' adaptation of locality principle, and therefore should be replaced by PI. The predictions of QM confirm PI and inherently refute OI~\cite{Shi1,Ebe}. So, it seems that QM is consistent with RC. Despite the extensive impact of Jarrett's project on Bell's literature, its implications are still a matter of controversy~\cite{Mau3,Nor3}. Norsen argued that the signalling requires some measure of {\it control} over appropriate beables on the part of the sender, and some measure of {\it access} on the part of the recipient. So, theories can exhibit violations of RC and yet (because certain beables are inadequately controllable by and/or accessible to humans) preclude superluminal signalling. Even if we admit Jarrett's analysis in the first place, we show that it is appropriate just in the first step of EPRB experiment, i.e., the preparation of the entangled state.\\
\subsubsection{Locality vs. Separability}
\indent In a more conceptual context, locality is supposed to be distinct from separability. This distinction is usually emphasised and traced back to Einstein by Howard. He argues that there are two logically independent, conceptually distinct principles he calls ``separability'' and ``locality''. The former, found in Einstein's own incompleteness argument~\cite{Ein}, implies that the state of a multi-component system is specified once the intrinsic state of each component is specified~\cite{How1,How2}. The latter requires that the physical state of a system is unaffected by events in regions of the universe so remote from the given system that no signal could connect them~\cite{How2}. The ``locality'' can be restored by dropping ``separability'' (a position sometimes referred to as ``holism''). He argues that ``separability'' implies OI and ``locality'' entails PI, if one supposes the individual probabilities as the marginals of a given joint probability, like~(\ref{Eq3}). This type of reasoning has been taken up by Shimony~\cite{Shi2} and Redhead~\cite{Red}, in which the denial of separability is claimed to block the derivation of the Bell's inequality. Developing on this theme, Healey has proposed an ``interactive'' interpretation of QM that relies heavily on non-separability, and discussed holism more broadly~\cite{Hea1,Hea2,Hea3,Hea4}. The separability program has been criticised on many grounds. Hensen has challenged the whole idea, claimed that neither separability is implied by Bell's locality condition, nor it is an implicit assumption of Bell's theorem~\cite{Hen}. The consequences of Howard's separability have been also criticized~\cite{Lau2,Mau3,Ber,Win,Fog}. Especially, Winsberg and Fine have argued that separability is not equivalent to OI~\cite{Win}. They accept Howard's conceptual definition of separability but cast doubt on equating determination with multiplication (OI). They assert that realizing separability requires just a function that maps the pairs of marginals to joints. In other words, OI is sufficient but not necessary for separability. Fogel shows that there is ambiguity in Winsberg and Fine's formulation of the subject and explores the various options~\cite{Fog}. Recently, De Raedt and co-workers emphasised that Einstein's concept of locality applies to individual events and therefore it doesn't necessarily lead to the probabilistic concept of Bell's locality condition~\cite{Rae}. Along these lines, Hess and Philipp argue that non-local theorems based on Bell's locality condition are not as general as they are claimed to be~\cite{Hes}. \\
\indent We begin with the very concept of separability in physics. Separability is generally assumed to be satisfied in isolated systems. A system is isolated if i) it is {\it dynamically} independent from any other system, i.e., there is no interaction Hamiltonian, and ii) it is {\it statistically} independent from any other system~\cite{Aul}. Note that, in QM, dynamical independence is only a necessary condition of an isolated system, while in Classical Mechanics they are equivalent. According to Howard, the separability is the principle of individuation for physical systems. Practically, this is equivalent to the first condition of an isolated system, i.e., dynamical independence. If we incorporate Howard's conceptual adaptation of separability into a realistic framework, the result is Bell's LC which, as we pointed out earlier, leads to Bell's locality condition through the measurement process. As a consequence, since ``locality'' is equivalent to PI, then ``separability'' results in ``locality'' through the measurement process. The ``locality'' is now closely related to Relativistic Causality. Maudlin seems to ignore this point and claims that ``locality'' is equivalent to LC~\cite{Mau3}.\\
\indent Howard's conceptual definition of separability is too general and thus cannot be applied to explicit physical theories. So, we should provide an explicit criterion for separability to carefully distinguish the problem of locality and that of separability. To do so, we address the problem of separability as it is entered into to the concept of Entanglement. An entangled state is a state of a multi-component system whose components are not statistically independent. So, entangled states are not separable. So, the statistical correlation can be a criterion for separability. The statistical correlation between two random variables is a scaled version of covariance. In QM, the covariance of observables $A$ and $B$ is defined as
\begin{equation}\label{Eq8}
Cov(A,B)=\langle AB\rangle-\langle A\rangle\langle B\rangle
\end{equation}
where $\langle\rangle$ denotes the expectation value. So, we make clear the notion of separability as
\begin{description}
   \item[Separability]{\it Two components $i$ and $j$ ($i\neq j$) are separable with respect to the physical properties $A^{(i)}$ and $B^{(j)}$, if their covariance is zero.}
\end{description}
Note that this is a {\it criterion}, not a {\it definition} for separability, and as a criterion, we do not intend to address it operationally, as in quantum information theory. It is consistent with the Howard's analysis. In an underlying theory, Howard's conceptual adaptation of separability leads to OI, which is equivalent to zero covariance. Our criterion, however, can be quantitatively applied to models in which assuming OI is not necessary. Moreover, since, at the underlying level, the statistical correlation is a function of marginals, it is consistent with Winsberg and Fine's formulation. Based on our criterion, one can show that the distinction between separability and locality is actually realizable~\cite{Sha}.
\subsubsection{Locality vs. Non-Contextuality}
It is believed that in the multi-component entangled systems, Bell's locality condition is equivalent to Non-Contextuality~\cite{Hom,Cli}. Non-Contextuality is ruled out by KS theorem~\cite{Koc}. The essence of KS theorem is that, for Hilbert spaces of dimension three or higher, the replacement of operators with their eigenvalues leads to the algebraically inconsistent relations. Traditionally, this means that the value assigned to each observable depends on the context in which it is measured, i.e, to the complete set of the compatible observables that are measured along with it. This description is only well-defined for QM. By contrast, Bell's locality condition applies to any physical theory that can be described operationally. Consequently, whereas one can examine whether the experimental statistics are consistent with Bell's locality condition (by examining whether they satisfy the inequalities), there is no way to examine whether Non-Contextuality is experimentally consistent. So, one should develop an operational criterion for it. Several attempts have been made to provide such a formulation~\cite{Cab,Sim,Lar}. More interestingly, Spekkens proposed a particularly natural generalization which applies to all models of any operational theory~\cite{Spe}. Let us first set out Spekkens's idea of the subject in our language.\\
\indent In QM, the primitives of description are preparations and measurements, specified as instructions for what should be done in the laboratory. The theory simply provides an algorithm for calculating the probability of an outcome of measurement, given a particular preparation. In a fundamental model of micro description, the preparation procedure includes a system in the state $\lambda$, and the measurement procedure is assumed to yield an outcome emerged from this state. So, we define non-contextuality condition as
\begin{description}
   \item[Non-Contextuality]{\it The underlying model of an operational theory is non-contextual, if two operationally-equivalent experimental procedures (including preparation and/or measurement) have the same representation in its formulation.}
\end{description}
Spekkens showed that it is not possible to embed QM into a preparation-based non-contextual model~\cite{Spe}. Moreover, he argues that non-locality within the framework of a separable model can be interpreted as measurement-based contextuality.\\
\indent Here, we present a simple approach to characterize the relationship between non-contextuality and locality through an extension of Mermin's approach on KS theorem~\cite{Mer}. Consider an ensemble of two half-spin entangled particles in the state $\phi$. Pauli spin operator for particle $i~(i=1,2)$ along the direction $j~(j=x,y)$ is represented by $\sigma_{ij}$. For such a system in QM, we have
\begin{align}\label{Eq9}
  \big[\sigma_{1x}\sigma_{2x},\sigma_{1y}\sigma_{2y}\big]&=0, \nonumber \\
  \big[\sigma_{1x}\sigma_{2y},\sigma_{1y}\sigma_{2x}\big]&=0
\end{align}
and
\begin{equation}\label{Eq10}
\sigma_{1x}\sigma_{2x}\sigma_{1y}\sigma_{2y}+\sigma_{1x}\sigma_{2y}\sigma_{1y}\sigma_{2x}=0
\end{equation}
In a non-contextual theory,~(\ref{Eq9}) leads to
\begin{align}\label{Eq11}
  \nu(\sigma_{1x}\sigma_{2x}\sigma_{1y}\sigma_{2y})&= \nu(\sigma_{1x})\nu(\sigma_{2x})\nu(\sigma_{1y})\nu(\sigma_{2y}), \nonumber \\
  \nu(\sigma_{1x}\sigma_{2y}\sigma_{1y}\sigma_{2x})&= \nu(\sigma_{1x})\nu(\sigma_{2y})\nu(\sigma_{1y})\nu(\sigma_{2x})
\end{align}
where $\nu(\sigma_{ij})$ represents the value of operator $\sigma_{ij}$. Also~(\ref{Eq10}) leads to
\begin{equation}\label{Eq12}
\nu(\sigma_{1x}\sigma_{2x}\sigma_{1y}\sigma_{2y})+\nu(\sigma_{1x}\sigma_{2y}\sigma_{1y}\sigma_{2x})=0
\end{equation}
Combining~(\ref{Eq11}) and~(\ref{Eq12}), one obtains
\begin{equation}\label{Eq13}
\nu(\sigma_{1x})\nu(\sigma_{2x})\nu(\sigma_{1y})\nu(\sigma_{2y})=0
\end{equation}
In a non-contextual theory, the value of each operator, independent of how it is measured, corresponds to one of its eigenvalues. Since the eigenvalues of Pauli spin operators are $\pm1$, the left hand side of~(\ref{Eq13}) should be $\pm1$, which is not. So,~(\ref{Eq13}) is denied. The origin of this disagreement might be traced back to the way~(\ref{Eq11}) is derived. It can be rewritten as
\begin{align}\label{Eq14}
  \nu(\sigma_{1x}\sigma_{2x}\sigma_{1y}\sigma_{2y})&=\nu(\sigma_{1x}\sigma_{2x})\nu(\sigma_{1y}\sigma_{2y}), \nonumber \\
  \nu(\sigma_{1x}\sigma_{2y}\sigma_{1y}\sigma_{2x})&=\nu(\sigma_{1x}\sigma_{2y})\nu(\sigma_{1y}\sigma_{2x})
\end{align}
and~(\ref{Eq12}) is then reconsidered as
\begin{equation}\label{Eq15}
\nu(\sigma_{1x}\sigma_{2x})\nu(\sigma_{1y}\sigma_{2y})+\nu(\sigma_{1x}\sigma_{2y})\nu(\sigma_{1y}\sigma_{2x})=0
\end{equation}
which leads to~(\ref{Eq13}) if
\begin{align}\label{Eq16}
  \nu(\sigma_{1x}\sigma_{2x})&=\nu(\sigma_{1x})\nu(\sigma_{2x}), \nonumber \\
  \nu(\sigma_{1y}\sigma_{2y})&=\nu(\sigma_{1y})\nu(\sigma_{2y}), \nonumber \\
  \nu(\sigma_{1x}\sigma_{2y})&=\nu(\sigma_{1x})\nu(\sigma_{2y}), \nonumber \\
  \nu(\sigma_{1y}\sigma_{2x})&=\nu(\sigma_{1y})\nu(\sigma_{2x})
\end{align}
Since~(\ref{Eq13}) is untenable,~(\ref{Eq16}) does not hold. This is because assigning values to the compatible operators $\sigma_{1x}\sigma_{2x}$ and $\sigma_{1y}\sigma_{2y}$ {\it in comparison with} the compatible operators $\sigma_{1x}\sigma_{2y}$ and $\sigma_{1y}\sigma_{2x}$ requires different preparations. Suppose that $\sigma_{1x}\sigma_{2y}$ and $\sigma_{1y}\sigma_{2x}$ are determined by a different state $\phi'$. Accordingly,~(\ref{Eq15}) can be rewritten as
\begin{equation}\label{Eq17}
\nu(\sigma_{1x}\sigma_{2x})\nu(\sigma_{1y}\sigma_{2y})+\nu'(\sigma_{1x}\sigma_{2y})\nu'(\sigma_{1y}\sigma_{2x})=0
\end{equation}
This is an appropriate alternate for~(\ref{Eq15}). Assuming the space-like separation of the particles and the independence of light cones of separated entities (as a consequence of~(\ref{Eq4})), one obtains
\begin{align}\label{Eq18}
\nu(\sigma_{1x})\nu(\sigma&_{2x})\nu(\sigma_{1y})\nu(\sigma_{2y})\nonumber \\
&+\nu'(\sigma_{1x})\nu'(\sigma_{2y})\nu'(\sigma_{1y})\nu'(\sigma_{2x})=0
\end{align}
which manifests a local contextual description. \\
\indent The foregoing discussion can be applied directly to EPRB experiment. Suppose that the spin of the particle $1$($2$) is measured along $a$($b$) or $a'$($b'$). The operators $\sigma_{1a(a')}\sigma_{2b(b')}$ are mutually incompatible for $a\neq a'$ and $b\neq b'$. So, the value of each operator depends on a different state preparation: $\nu_{p}(\sigma_{1a(a')}\sigma_{2b(b')})$, where the subscript $p$ refers to the state preparation. The description is preparation-based non-contextual, if one reads
\begin{equation}\label{Eq19}
\nu_{p}(\sigma_{1a(a')}\sigma_{2b(b')})=\nu(\sigma_{1a(a')}\sigma_{2b(b')})
\end{equation}
Due to the fact that non-locality can be interpreted as measurement contextuality, the description is not yet explicitly non-contextual. It would be so, if we apply Bell's locality condition on values
\begin{equation}\label{Eq20}
\nu(\sigma_{1a(a')}\sigma_{2b(b')})=\nu(\sigma_{1a(a')})\nu(\sigma_{2b(b')})
\end{equation}
Now, one can derive CHSH inequality (see e.g.,\cite{Brau}). Assuming Bell's locality condition alone,
\begin{equation}\label{Eq21}
\nu_{p}(\sigma_{1a(a')}\sigma_{2b(b')})=\nu_{p}(\sigma_{1a(a')})\nu_{p}(\sigma_{2b(b')})
\end{equation}
no Bell-type inequalities can be concluded. In the other words, Bell's locality condition is not sufficient for non-contextuality, a point that manifests itself in the single-particle entangled systems~\cite{Hom,Cli,ShGo}.\\
\indent According to KS theorem, QM is a contextual theory (for an opposing perspective, see~\cite{Gri}). Our approach shows that, at the fundamental level, due to the fact that Bell's locality condition can be considered as a consequence of non-contextuality, it is impossible to construct a non-local, non-contextual underlying theory.
\section{3. A New Examination of Three Steps of EPRB Experiment}
Up to now, EPRB experiment has been examined by different approaches in the literature. All these approaches describe only the initial entangled state of the system. There are, however, two measurement processes, each performed on one particle, which should be examined as well. Here, based on proposed criteria, we describe the events occurred in three steps of EPRB experiment by both QM and HV theories.
\subsection{I. State Preparation}
In this step, a source emits two entangled particles, prepared in a singlet state
\begin{equation}\label{Eq22}
|\psi_{i}\rangle=\frac{1}{\sqrt{2}}\big(|a,+\rangle_{1}|a,-\rangle_{2}-|a,-\rangle_{1}|a,+\rangle_{2}\big)
\end{equation}
where $|a,+\rangle_{1}$ represents the eigenstate of Pauli spin component of particle $1$ along the direction $a$ with eigenvalue $+1$ and so on. The joint probability for the outcomes $A$ and $B$ for operators $\sigma_{1a}$ and $\sigma_{2b}$ is obtained by
\begin{equation}\label{Eq23}
P(A,B|a,b,\psi_{i})=\frac{1}{4}(1-AB\cos{\theta_{ab}})
\end{equation}
where $\theta_{ab}=|a-b|$. The marginal probability for each particle is obtained from this joint probability as $\frac{1}{2}$, which shows that outcomes of measurements on individual particles are completely random. The conditional probability of the outcome $B$ for operator $\sigma_{2b}$, given the outcome $A$ for operator $\sigma_{1a}$, would be
\begin{equation}\label{Eq24}
P(B|a,b,A,\psi_{i})=\frac{1}{2}(1-AB\cos{\theta_{ab}})
\end{equation}
The difference between this conditional probability and the corresponding marginal probability confirms the existence of the correlation. Consequently, one can easily show that
\begin{equation}\label{Eq25}
Cov(\sigma_{1a},\sigma_{2b})=-\cos{\theta_{ab}}
\end{equation}
So, at the quantum level, two particles are not separable. On the other hand, since the probability distributions depend on the initial entangled state of the system, the description is contextual. The measurement has not been performed yet, so Bell's locality condition is not involved.\\
\indent Let us explain the underlying descriptions of this step. In a HV contextual theory, $\lambda$ includes the state preparation of the system. Suppose $\sigma_{1a}$ is measured along with $\sigma_{2b}$ or $\sigma_{2b'}$. Since $[\sigma_{1a},\sigma_{2b(b')}]=0$ and $[\sigma_{2b},\sigma_{2b'}]\neq0$, the state of the system is determined by $\lambda^{a}_{b}$ or $\lambda^{a}_{b'}$, corresponding to the probabilities $P(A,B|a,b,\lambda^{a}_{b})$ or $P(A,B|a,b',\lambda^{a}_{b'})$. Suppose that the probabilities of the outcomes are determined independent of the state preparation of the system. If so, the probability of a certain outcome of $\sigma_{1a}$ should be the same for both state preparations. However, this is not sufficient to realize non-contextuality, because the relationship between $\sigma_{1a}$ and $\sigma_{2b}$ ($\sigma_{2b'}$) has not yet been determined. Non-contextuality is fulfilled, if Bell's locality condition is introduced. Bell's locality condition is the logical conjunction of OI and PI. It has been shown that a theory that accounts for the predictions of QM, should include some non-local features, either by a violation of PI or by a violation of OI~\cite{Von}. We consider two possible such theories: the one that respects PI and violates OI, and the one that does the opposite.\\
\indent Let us start with a HV theory that violates PI and respects OI. If we incorporate OI into Bayes' theorem~(\ref{Eq3}), we obtain
\begin{equation}\label{Eq26}
P(A,B|a,b,\lambda)=P(A|a,b,\lambda)P(B|a,b,\lambda)
\end{equation}
Note that since non-locality due to the violation of PI can be interpreted as contextuality, the description is still contextual. The average value of the spin component of each particle is obtained as
\begin{align}\label{Eq27}
  E^{(1)}(a,b,\lambda)&=\sum_{A}A~P(A|a,b,\lambda)\nonumber \\
E^{(2)}(a,b,\lambda)&=\sum_{B}B~P(B|a,b,\lambda)
\end{align}
and the joint average value would be
\begin{align}\label{Eq28}
E^{(12)}(a,b,\lambda)&=\sum_{A,B}AB~P(A,B|a,b,\lambda)\nonumber \\
&=E^{(1)}(a,b,\lambda)E^{(2)}(a,b,\lambda)
\end{align}
Accordingly,
\begin{equation}\label{Eq29}
Cov(\sigma_{1a},\sigma_{2b})=0
\end{equation}
So, a HV theory based on OI satisfies separability. If we integrate over $\lambda$, we arrive at the following average values which should be equal to the expectation values of QM
\begin{align}\label{Eq30}
  \langle\sigma_{1a}\rangle_{b}&=\int_{\Lambda} d\lambda~\rho(\lambda)E^{(1)}(a,b,\lambda)\nonumber \\
  \langle\sigma_{2b}\rangle_{a}&=\int_{\Lambda} d\lambda~\rho(\lambda)E^{(2)}(a,b,\lambda)
\end{align}
and
\begin{equation}\label{Eq31}
\langle\sigma_{1a}\sigma_{2b}\rangle=\int_{\Lambda} d\lambda~\rho(\lambda)E^{(12)}(a,b,\lambda)
\end{equation}
The subscript $b$ in $\langle\sigma_{1a}\rangle_{b}$ (the subscript $a$ in $\langle\sigma_{2b}\rangle_{a}$) shows the contingency of outcomes of the first (second) particle on the measurement setting of the second (first) particle. Note that since in this step $\langle\sigma_{1a}\rangle_{b}=\langle\sigma_{2b}\rangle_{a}=0$, this contingency is not experimentally conceivable.\\
\indent In conclusion, it is not still clear whether a theory based on the violation of PI is able to reproduce the predictions of QM or not. We answer to this question in the next step of EPRB experiment. \\
\indent Now, we consider a model theory that respects PI and violates OI. Suppose that the probability distribution of the second particle depends on the outcomes of the first particle
\begin{equation}\label{Eq32}
P(A,B|a,b,\lambda)=P(A|a,\lambda)P(B|a,b,A,\lambda)
\end{equation}
Note that the violation of PI for the second particle arises from the violation of OI. It is believed that since the non-locality due to the violation of OI is uncontrollable, it exposes no conflict with RC. This type of non-locality is recognized by Shimony as ``passion at a distance''~\cite{Shi1} and by Gisin and Go as ``quantum non-locality''~\cite{GiG}. The violation of OI results in a non-separable description
\begin{equation}\label{Eq33}
E^{(12)}(a,b,\lambda)\neq E^{(1)}(a,\lambda)E^{(2)}(a,b,\lambda)
\end{equation}
Accordingly, non-locality due to the violation of OI cannot necessarily be interpreted as the measurement-based contextuality. The marginal probabilities are obtained from the joint probability~(\ref{Eq32}) as
\begin{align}\label{Eq34}
  P(A|a,\lambda)&=\sum_{B}P(A,B|a,b,\lambda) \nonumber \\
  P(B|a,b,A,\lambda)&=\sum_{A}P(A,B|a,b,\lambda)
\end{align}
The average value of the spin component of each particle is obtained as
\begin{align}\label{Eq35}
  E^{(1)}(a,\lambda)&=\sum_{A}A~P(A|a,\lambda)\nonumber \\
E^{(2)}(a,b,\lambda)&=\sum_{B}B~P(B|a,b,A,\lambda)
\end{align}
and the joint average value would be
\begin{equation}\label{Eq36}
E^{(12)}(a,b,\lambda)=\sum_{A,B}AB~P(A,B|a,b,\lambda)
\end{equation}
If we integrate over $\lambda$, we arrive at the expectation values of QM as
\begin{align}\label{Eq37}
  \langle\sigma_{1a}\rangle&=\int_{\Lambda} d\lambda~\rho(\lambda)E^{(1)}(a,\lambda)\nonumber \\
  \langle\sigma_{2b}\rangle_{a}&=\int_{\Lambda} d\lambda~\rho(\lambda)E^{(2)}(a,b,\lambda)
\end{align}
and
\begin{equation}\label{Eq38}
\langle\sigma_{1a}\sigma_{2b}\rangle=\int_{\Lambda} d\lambda~\rho(\lambda)E^{(12)}(a,b,\lambda)
\end{equation}
There is not any no-go theorem excluding the possibility of a theory based on the violation of OI. So, such a theory might be a good candidate theory to reproduce the predictions of QM. We elaborate this issue more in the second step of EPRB experiment.\\
\indent By applying Bell's locality condition, Bell's theorem is proved considering a HV theory which is separable, potentially local and non-contextual. Although Bell's theorem prohibits non-contextual theories, it doesn't clarify what kind of a HV theory is able to reproduce the predictions of QM. We examine this question in the next step of EPRB experiment.
\subsection{II. Measurement on the First Particle}
In this step, $\sigma_{1a}$ is measured and suppose that the outcome $A'$ is obtained. Accordingly, the state of the system is reduced to
\begin{equation}\label{Eq39}
|\psi_{m}\rangle=|a,A'\rangle\otimes\frac{1}{2}\sum_{B=\pm A'}\Big(-Be^{\frac{\textrm{i}\theta_{ab}}{2}}+e^{-\frac{\textrm{i}\theta_{ab}}{2}}\Big)|b,B\rangle
\end{equation}
Then, the probability of outcome $B$ for the second particle would be
\begin{equation}\label{Eq40}
P(B|a,b,\psi_{m})=\frac{1}{2}\big(1-A'B\cos{\theta_{ab}}\big)
\end{equation}
It is obvious that
\begin{align}\label{Eq41}
  \langle\sigma_{1a}\rangle&=A' \nonumber \\
  \langle\sigma_{2b}\rangle_{a}&=\sum_{B}B~P(B|a,b,\psi_{m})=-A'\cos{\theta_{ab}}
\end{align}
The dependency of the probability distribution of the second particle on the constant $A'$, and the angle $\theta_{ab}$ is traced back to the state preparation of the system. Accordingly, the quantum description in this step is preparation-based contextual. Also, since $\langle\sigma_{1a}\sigma_{2b}\rangle=\langle\sigma_{1a}\rangle\langle\sigma_{2b}\rangle_{a}=-\cos{\theta_{ab}}$, two particles are separable
\begin{equation}\label{Eq42}
Cov(\sigma_{1a},\sigma_{2b})=0
\end{equation}
It should be noticed that the expectation value $\langle\sigma_{1a}\sigma_{2b}\rangle$ is the same as the first step. Since the measurement is performed in this step, locality is involved. The probability distribution~(\ref{Eq41}) shows that, contrary to the first step, the quantum description in the second step, refutes PI and satisfies OI. This can be also checked out by separability criterion~(\ref{Eq42}), because OI results in separability. Note that the outcome $A'$ is a constant and should not be considered as the violation of OI. So, it seems that the program of ``peaceful coexistence'' between QM and RC is untenable in the second step. Since the description is separable, the non-locality due to the violation of PI can be interpreted as a measurement-based contextuality.\\
\indent Now, we consider the underlying descriptions of the second step. In a non-contextual HV theory based on Bell's locality condition~(\ref{Eq4}), the average value $E^{(2)}(b,\lambda)$ is independent of the measurement setting $a$. So, the dependency of $\langle\sigma_{2b}\rangle$ on $a$ in the second step cannot be resulted from it, i.e.,
\begin{equation}\label{Eq43}
\langle\sigma_{2b}\rangle_{a}^{2nd}\neq\int_{\Lambda}d\lambda~E^{(2)}(b,\lambda)
\end{equation}
Accordingly, such a theory is not consistent with the predictions of QM in both steps. The impossibility of non-contextual theories can be directly interpreted from Bell's theorem, which we presented it here by a different line of reasoning. A similar objection is raised for a contextual theory based on the violation of PI, in which contextuality, independent of the state preparation, is considered due to the measurements performed on the first particle. Then, averaging on $E^{(1)}(a,b,\lambda)$ (and also $E^{(2)}(a,b,\lambda)$) over $\lambda$ should be equal to zero in the second step, because it is zero in the first step. Therefore, such a theory cannot be consistent with QM either.\\
\indent In the last section, we showed that an underlying theory based on the violation of OI might be a candidate model consistent with QM. In the second step, since the measurement of $\sigma_{1a}$ yields the outcome $A'$, then
\begin{equation}\label{Eq44}
P(A|a,\lambda)=\delta_{A,A'}
\end{equation}
Accordingly, one gets
\begin{align} \label{Eq45}
E^{(1)}(a,\lambda)&=\sum_{A}A~P(A|a,\lambda)=A'\nonumber \\
E^{(2)}(a,b,\lambda)&=\sum_{B}B~P(B|A',a,b,\lambda)
\end{align}
Note that if we introduce the definite outcome $A'$, the marginal probability of the second particle in the second step would be equal to its conditional probability in the first step. If we integrate over $\lambda$, we arrive at the expectation values of QM
\begin{align}\label{Eq46}
  \langle\sigma_{1a}\rangle&=A'\nonumber \\
  \langle\sigma_{2b}\rangle_{a}&=\int_{\Lambda} d\lambda~\rho(\lambda)E^{(2)}(a,b,\lambda)
\end{align}
So, in the second step, the underlying theory based on the violation of OI can be consistent with QM.
\subsection{III. Measurement on the Second Particle}
In this step, $\sigma_{2b}$ is measured and suppose that the outcome $B'$ is obtained. Accordingly, the state of the system is reduced to
\begin{equation}\label{Eq47}
|\psi_{f}\rangle=|a,A'\rangle_{1}\otimes|b,B'\rangle_{2}
\end{equation}
Since $\langle\sigma_{1a}\rangle=A'$, $\langle\sigma_{2b}\rangle=B'$ and $\langle\sigma_{1a}\sigma_{2b}\rangle=A'B'$, the quantum description is separable ($Cov(\sigma_{1a},\sigma_{2b})=0$). On the other hand, due to the fact that the outcome of the second particle is determined by the reduction of the state of the second particle to $|b,B'\rangle_{2}$, the description is preparation-based non-contextual. Since Bell's locality condition is satisfied, the description is also measurement-based non-contextual.\\
\indent Let us examine this step at the underlying level. In an underlying contextual theory based on the violation of OI, we have
\begin{equation}\label{Eq48}
P(B|a,b,A,\lambda)=\delta_{B,B'}
\end{equation}
It is obvious that from the first to the third step, the information about the states of the particles gradually increases (see~(\ref{Eq44}) and~(\ref{Eq48})). To explore the details of the issue in an explicit model see~\cite{Sha1}.
\section{IV. Conclusion}
It is commonly believed that Bell's theorem excludes a local realistic reconstruction of experimentally-verified predictions of QM regarding EPRB thought experiment. The question is that what kind of the theory can exactly be consistent with QM? To address this question, it is important to make a detailed examination of the role of a crucial assumption underlying the theorem; Locality. Depending on the context, there are different adaptations of this assumption. The usual suspects are Local Causality, Factorizability, Relativistic Causality, Separability and Non-Contextuality. Although they are overlapping, each relies on a specific aspect of the subject. Factorizability, better known as Bell's locality condition is the logical conjunction of PI and OI, concluded from Bell's own idea of locality, known as Local Causality. It is believed that Relativistic Causality is equivalent to  impossibility of superluminal signalling. Contrary to OI, non-locality due to the violation of PI can be used for signalling. QM respects PI and violates OI. So, it seems that QM is consistent with Relativistic Causality. We showed that this is not the whole picture. Actually, this is the case just in the first step of EPRB experiment, i.e., the preparation of the initial entangled state. There are, however, two measurement processes, each performed on one of the particles. In the second step, including the measurement on the first particle, the quantum description, according to~(\ref{Eq41}), violates PI. So, Relativistic causality. In the third step, including the measurement on the second particle, the quantum description respects both OI and PI. On the other hand, we argued that Separability is equivalent to Local Causality, which leads to Bell's locality condition through the measurement process. The constraint of zero covariance is sufficient to define separability. We defined the independence of representations from the preparation and measurement contexts, as Non-Contextuality. The preparation-based non-contextuality is an {\it independent} assumption, but non-locality within the framework of a {\it separable} model can be interpreted as measurement-based contextuality. According to the introduced analysis, we showed that the following logical relations is concluded
\begin{align}
   PI\wedge OI&\Longrightarrow Separability\wedge\neg Contextuality  \label{Eq49}\\
   \neg PI\wedge OI&\Longrightarrow Separability\wedge\neg Locality  \label{Eq50}\\
   \neg OI&\Longrightarrow \neg Separability  \label{Eq51}
\end{align}
While Bell's theorem refutes Bell's locality condition in~(\ref{Eq49}), our approach (which is more resonant than Leggett's theorem) shows that an underlying theory based on the violation of PI in~(\ref{Eq50}) is also inconsistent with QM. In effect, only an underlying theory based on the violation of OI in~(\ref{Eq51}) is capable of reproducing the predictions of QM in the EPRB experiment. Incidentally, a recent event-by-event computer model based on the violation of OI that exploits the coincidence-time loophole has developed and reproduced the predictions of QM for EPRB experiment~\cite{Rae}.

\end{document}